\documentclass[reprint]{revtex4-1}

\usepackage[english]{babel}
\usepackage{braket}
\usepackage{epsfig}
\usepackage{graphics}
\usepackage{graphicx}
\usepackage{xcolor}
\usepackage{hyperref}
\usepackage{natbib}
\usepackage{amsmath}
\usepackage{times}
\usepackage[T1]{fontenc}
\usepackage[utf8]{inputenc}
\usepackage{ulem}
\usepackage{psfrag}
\usepackage{epstopdf}

\usepackage[toc]{appendix}

\begin{document}

\title{Quantifying wave
propagation in a chain of FitzHugh-Nagumo neurons}

\author{\textbf{L. Messee Goulefack}}
\email[messeelinda20@gmail.com]{(Corresponding author)}
\affiliation{Fundamental Physics Laboratory, Department of Physics, Faculty of Science, University of Douala, Box 24157 Douala, Cameroon}
\affiliation{Department of Physics, Pontifical Catholic University of Rio de Janeiro, Rua Marqu\^es de S$\tilde{a}$o Vicente, 225--22451--900 G\'avea - Rio de Janeiro - RJ Brazil}

\author{\textbf{C. Masoller }}
 \affiliation{Department of Physics, Universitat Politecnica de Catalunya, Rambla St. Nebridi 22, Terrassa 08222, Barcelona, Spain }

\author{\textbf{R. Yamapi}}
 \affiliation{Fundamental Physics Laboratory, Department of Physics, Faculty of Science, University of Douala, Box 24157 Douala, Cameroon}

 \author{\textbf{C. Anteneodo}}
\affiliation{Department of Physics, Pontifical Catholic University of Rio de Janeiro, Rua Marqu\^es de S$\tilde{a}$o Vicente, 225--22451--900 G\'avea - Rio de Janeiro - RJ Brazil}
\affiliation{ National Institute of Science and Technology (INCT) of Complex Systems, Rua Marqu\^es de S$\tilde{a}$o Vicente, 225--22451--900 G\'avea - Rio de Janeiro -Brazil }


\begin{abstract}
Understanding how external stimuli propagate in neural systems is an important challenge in the fields of neuroscience and nonlinear dynamics. 
Despite extensive studies over several decades, this problem remains poorly understood. In this work,  we examine a simple ``toy model'' of an excitable medium, a linear chain of diffusely coupled FitzHugh-Nagumo neurons, and analyze the transmission of a sinusoidal signal injected into one of the neurons at the ends of the chain.
We measure to what extent the propagation of the wave reaching the opposite end is affected by the frequency and amplitude of the signal, the number of neurons in the chain and the strength of their mutual diffusive coupling.   
To quantify these effects, we measure the cross-correlation between the time-series of the membrane potentials of the end neurons.
This measure allows us to detect the values of the parameters that delimit different propagation regimes.


\end{abstract}
\maketitle

\section{Introduction}
 \label{sec:introduction}
 
Neuronal models are crucial in the exploration and understanding of oscillation modes of neural systems\cite{gallas,gallas2}, serving as foundational tools to investigate both normal and pathological neuronal states. The FitzHugh-Nagumo (FHN) model~\cite{FitzHugh1961,Nagumo1962} has been extensively employed to study neuronal excitability and the dynamic behavior of single neurons and neuronal ensembles. Its ability to capture essential features of neuronal firing, while remaining computationally efficient, makes it an ideal model for examining large-scale neural dynamics, including synchronization, wave propagation, and pattern formation~\cite{Izhikevich2007,keener2009}.

The coupling of FHN neurons through diffusive interactions has been a subject of intense study, providing insights into how local interactions can lead to complex global behaviors such as synchronization and desynchronization~\cite{Haken2004,Bressloff2015}. Diffusive coupling, which mimics the electrical interactions between adjacent neurons, is fundamental in coordinating neuronal activity. A coordination which is critical in various brain functions, such as sensory processing, motor control, and cognitive tasks ~\cite{Destexhe2004, Kumar2010}.

External inputs stimulate neuronal dynamics, influencing both physiological and pathological states. 
Local stimulation can induce long-range order~\cite{baier1999local} and transient coherence~\cite{baier2000human}.
Therefore, it is relevant to study how an external stimulus propagates and affects the neural dynamics. In this respect, due to their periodic and elementary shape,  
sinusoidal signals have been used to study the response properties of single neurons and neuronal ensembles~\cite{Glass1988,longin_1991,longtin_1998,chacron_2000,baier2004,reinoso_2016}. These studies reveal that neuronal systems can exhibit resonance phenomena, where certain frequencies of external inputs maximize or optimize  the system's response.  
Studies have also explored how resonant response to external stimuli might modulate or disrupt pathological synchronization~\cite{Buzsaki2004, Tass2003}.
In addition to neuronal models, biochemical oscillator networks have also been used as excitable media where local perturbations can be enhanced~\cite{baier1999local,baier2000human}. 
In such systems, it was found that an external periodic perturbation of an  oscillator produced, for certain ranges of the stimulation frequency, the appearance of globally coherent states. 
 
However, how external signals propagate in neural systems is still far from understood, and the purpose of the present work is to analyze signal propagation in a linear chain of diffusely coupled FHN neurons, considering sinusoidal and noisy signals. 
Our study contributes to ongoing efforts to better understand how external inputs influence neural behavior and timing. 

The manuscript is organized as follows. 
In Sect.~\ref{sec:model}, we describe the excitable media model used and signal injection. 
In Sect.~\ref{sec:correlation}, we define the cross-correlation quantifier $C$, between the membrane potential of the end neurons, used for the detection of signal propagation. The results are presented in Sect.~\ref{sec:results}, where we vary signal and network parameters, and 
discuss the different propagation regimes detected by the cross-correlation. 
We summarize the main findings and discuss future perspectives in Sect.~\ref{sec:results}.

\section{Model}
 \label{sec:model}

We analyze the response of a one-dimensional excitable medium to an applied sinusoidal signal, using a model that was previously studied in Refs.~\cite{baier2004,chernihovskyi2005}. The excitable medium consists of a  chain of $N$ diffusively-coupled identical FHN neurons and the sinusoidal signal is injected into one of the neurons at the ends of the chain. Governing equations are 
\begin{align} \notag
\frac{dx_i}{\epsilon dt}&=x_i(a-x_i)(x_i-1)- y_i + I_{ext} + D_i + \theta_i (t)/\epsilon,\\
\label{eq:network}
 \frac{dy_i}{\epsilon dt}&=bx_i-c y_i,
\end{align}
where $x_i$ is the membrane potential and $y_i$ is the recovery current of the $i$th neuron. $I_{ext}$ is the external stimulus current, and $a$, $b$ and $c$ are positive parameters, which we keep fixed as $(a,b,c,I_{ext})=(0.1, 0.015,0.015,0.062)$.  
We choose these parameter values because they place the single neuron dynamics just above the critical threshold, enhancing the sensitivity of the neurons to excitation inputs. 
$\epsilon$ denotes a time scaling coefficient, which allows to  adjust the resonant frequency of the neurons. Finally $D_i$ are diffusive coupling functions defined with zero-flux boundary conditions:
\begin{align*}
	D_1&=D_x(x_{2}-x_1),\\
D_i&=D_x(x_{i+1}-2x_i+x_{i-1})	\quad \text{for}\quad i=2, \; 3\;, \ldots \; N-1,\\
	D_N&=D_x(x_{N-1}-x_{N}) ,
\end{align*}
with coupling strength $D_x$. Diffusive coupling means that the activity of one neuron is influenced by the difference between its activity and that of its nearest neighbors. We select $D_x$ to be large enough to produce local excitation but not long-range order~\cite{baier2004}.
Throughout our study, we use $D_x=0.04$ for networks of size $N=20$, 
and set $\epsilon=10$, following 
Refs.~\cite{baier2004,chernihovskyi2005} to allow  comparisons. 
  Additionally, the neurons are subjected to external time-dependent inputs $\theta_i(t)$ with zero mean. 
To study the propagation of a single external input, 
from an injection time $t_{in}$, 
 a sinusoidal signal of amplitude $A$ and angular frequency $w$ is added to the first neuron. Therefore,
 \begin{eqnarray} \label{eq:theta}
     \theta_1(t) &=& A\sin( {w} t),\;\;\; \forall t>t_{in}, \\
     \theta_i(t) &=& 0,\;\;\; \forall t, \forall i\ne 1.
 \end{eqnarray}

\begin{figure}[h!] 
\includegraphics[width=0.5\textwidth]{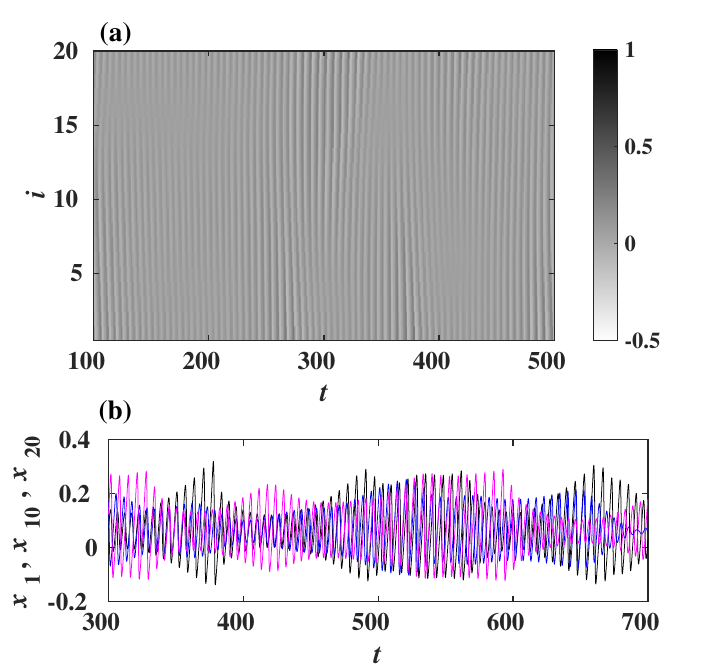}
\caption{ 
(a) Raster plot of the membrane potentials, $x_i(t)$, $i=1,\ldots N$, in gray scale, for a chain on $N=20$ FHN neurons, without  external input ($A = 0.0$).
In this figure and the following ones, the grayscale represents the value of the membrane potential, with darker shades indicating higher values, as defined by the bar on the right.  
(b) Time-series of the membrane potential of three neurons, $x_1$, $x_{10}$ and $x_{20}$. The maximum of the cross-correlation between $x_1$ and $x_{20}$, estimated over 100 realizations with random initial conditions is $C_{max}=0.61\pm 0.14$.
}
\label{fig:spaciow0}
\end{figure}
The model equations were integrated using a standard 4th order Runge-Kutta algorithm with $dt=10^{-2}$.   
A typical  raster  plot obtained for $N=20$ neurons, when no  external input is applied, is presented in Fig.~\ref{fig:spaciow0}. We can observe that for the coupling strength considered, the neurons do not oscillate regularly and are only partially synchronized.

\section{Quantifying signal propagation}
\label{sec:correlation}

To quantify the wave propagation induced when a signal is injected, we measure  the cross-correlation between the time series of the first and last neurons in the chain. This  is an effective procedure to assess whether the signal injected into the first  neuron reaches the last and to quantify their similarity  as a function of the time lag $\tau$, defined as 
\begin{equation}
\label{eq:crosscorelation}
   C\equiv C_{x_1x_N}(\tau) =  
    \frac{\langle ( x_1(t) - \langle{x}_1\rangle)(   x_N(t+\tau) - \langle{x}_N\rangle) \rangle}{ \sigma_{x_1} \sigma_{x_N} } ,
\end{equation}
where $\langle \cdots \rangle$ and $\sigma_x$ denote the average of $x$ and its standard deviation, respectively. We calculate these quantities in an observation window whose width, $\Delta t=200$, is selected to contain several periods of the input signal, for the entire frequency range studied. The window typically starts at $t_0=800$ (i.e., we analyze the time interval [800,1000]) in order to wait a time that is long enough (i.e., not too premature), for the signal to reach the opposite end. The role of the waiting time as a function of the number of neurons in the chain is discussed at the end of the next section.

\begin{figure*}[t!] 
\includegraphics[width=0.4\textwidth]{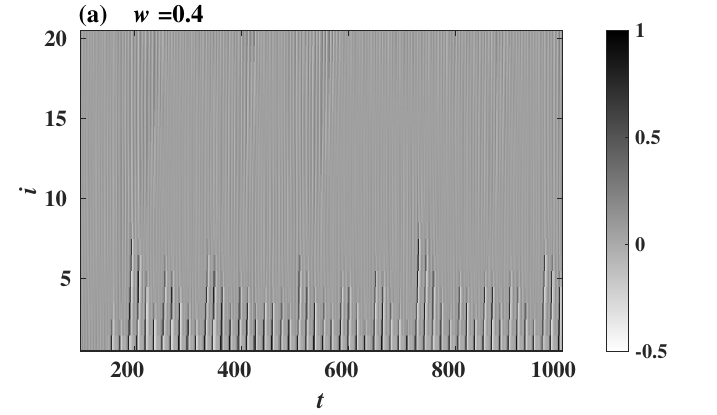}  
\includegraphics[width=0.4\textwidth]{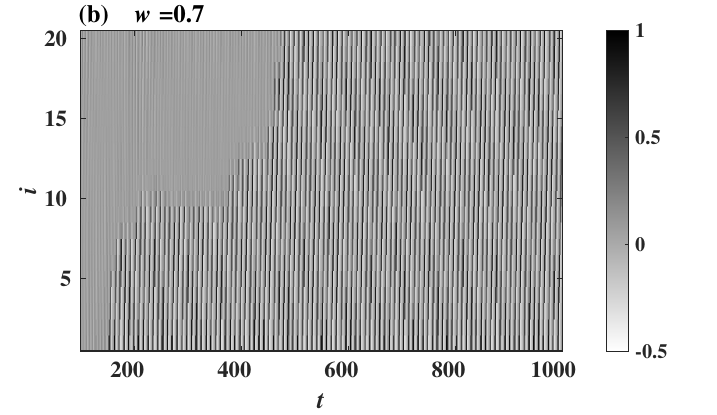} \\
\includegraphics[width=0.4\textwidth]{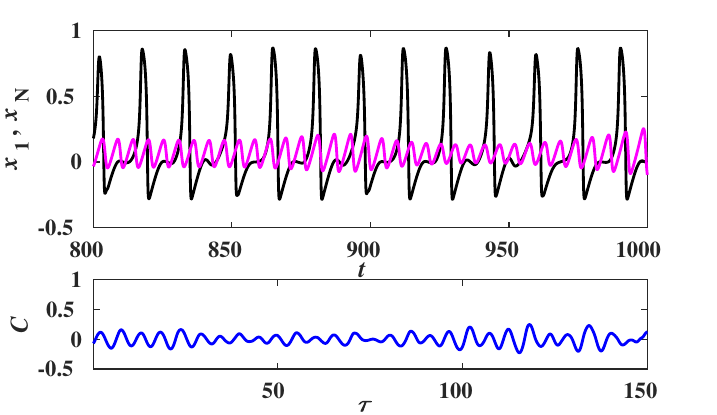} 
\includegraphics[width=0.4\textwidth]{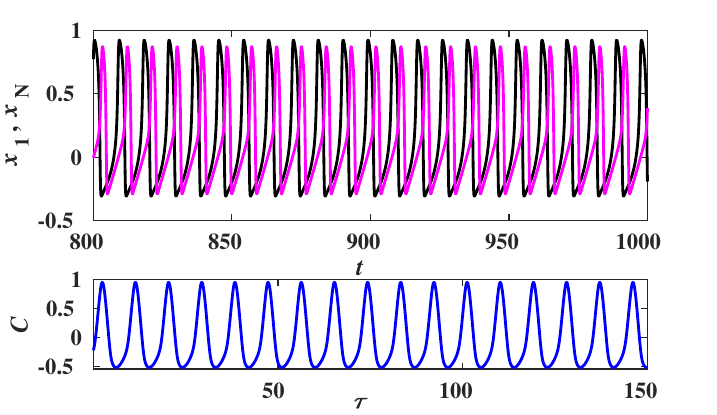} \\
\includegraphics[width=0.4\textwidth]{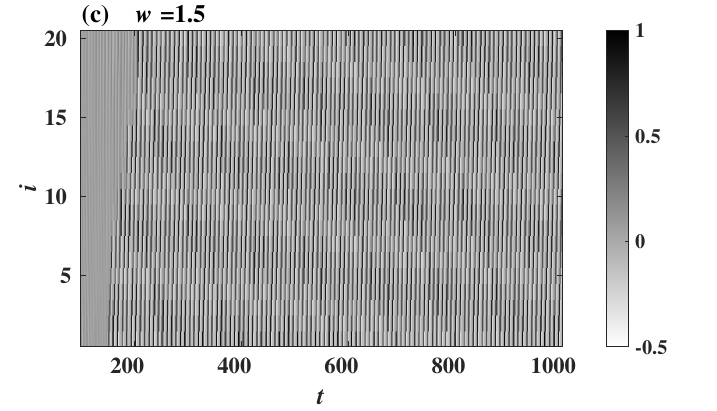}   
\includegraphics[width=0.4\textwidth]{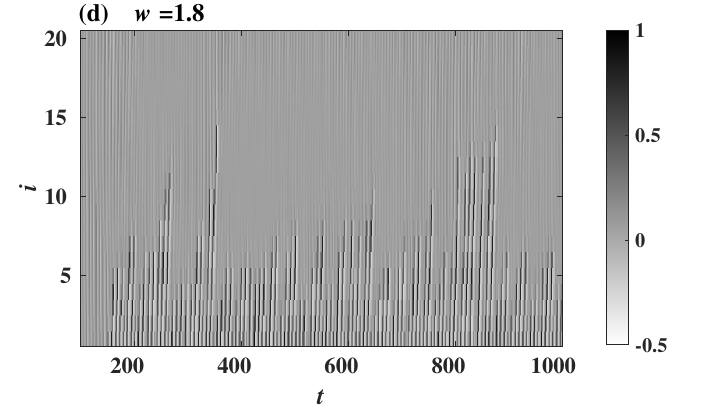}  \\
\includegraphics[width=0.4\textwidth]{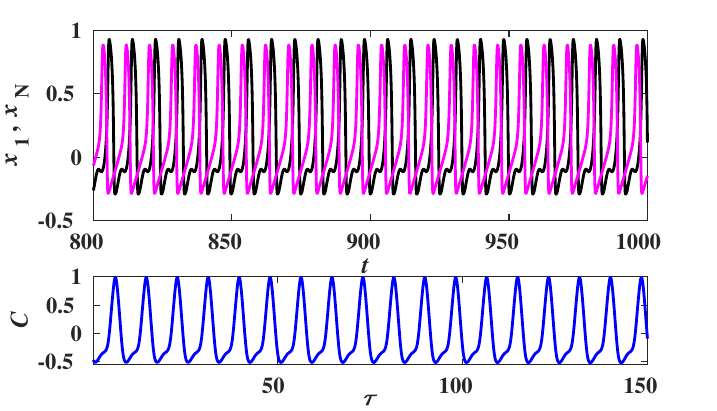}
\includegraphics[width=0.4\textwidth]{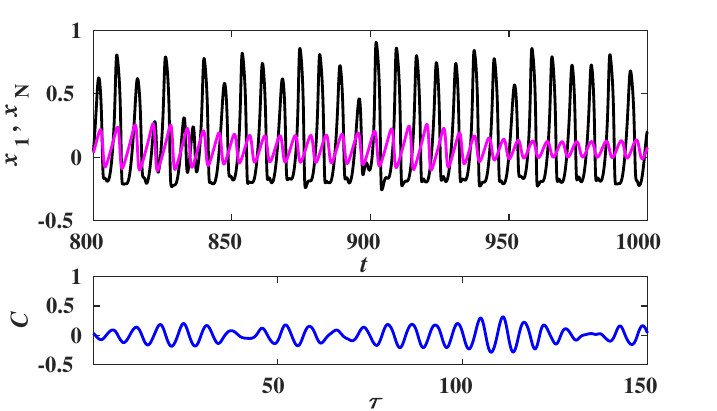}  
\caption{ Raster plots of the neurons' oscillations (top panels: $x_i(t)$, $i=1,\ldots N$ in gray scale), when the linear chain has $N=20$ neurons, for different values of the signal's angular frequency, $ {w}=$ 0.4 (a), 0.7 (b), 1.5 (c) and 1.8 (d). The signal's amplitude is $A = 0.3$ and the coupling strength is $D_x = 0.04$. We also present the time evolution of  neurons 
 located at the extremes (central panels -- $x_1$ in black and $x_N$ in magenta) and the cross-correlation between them vs. the time lag $\tau$ (lower panels), measured in the time interval $t \in [800,1000]$. We observe that in the time interval considered, when $ {w}=$ 0.4 and 1.8 the signal does not reach to the end of the chain, while when $ {w}=$ 0.7 and 1.5 the signal reaches the end neuron.
}
\label{fig:space-time-w20}
\end{figure*}

\section{Results}
\label{sec:results}

In this section we perform a systematic quantitative study of the effects of the frequency and amplitude of the sinusoidal signal added to the first neuron; we also discuss the effects of additive white noise. In all cases we use the cross-correlation to detect the values of the parameters that define the regimes where propagation occurs.

 \subsection{Effect of the signal's frequency}

In  Fig.~\ref{fig:space-time-w20}, 
we plot the  raster diagrams for a typical network where we injected, from time $t_{in}=150$ until the end of the run, the signal $\theta_1(t)$ defined in Eq.~(\ref{eq:theta}), considering different values of the angular frequency $w$.

For some frequencies the signal resonates with the neurons' oscillations, reaching the end of the chain, while for other frequencies the signal does not propagate far enough to reach the opposite end, as already known from Ref.~\cite{baier2004}. 
This can be noticed more clearly in the central plots where we display the time series $x_1(t)$ (dark black) and $x_2(t)$ (light magenta). 
Similar outcomes have been observed when, instead of neurons, the units in the chain model calcium oscillations~\cite{baier1999local,baier2000human}. 

When propagation occurs, the neurons' dynamics becomes periodic and the spiking activity of the two neurons at the two ends of the linear chain is very similar, as it can be seen when comparing 
$x_1$ and $x_N$ in Figs.~\ref{fig:space-time-w20}(b) and \ref{fig:space-time-w20}(c). 
Furthermore, larger frequencies allow faster propagation, denoted by the larger slope of the fronts in the  raster  plots (this can be seen when comparing the  raster  plot for $ {w}=1.5$ shown in Fig.~\ref{fig:space-time-w20}(c) with that for $ {w}=0.7$, shown in Fig.~\ref{fig:space-time-w20}(b)).

Now, to quantify the degree of propagation and detect the values of the parameters that delimit different propagation regimes, we use the cross-correlation $C$  between the membrane potentials of the end neurons, defined in Eq.~(\ref{eq:crosscorelation}), that we plot as a function of the time lag $\tau$ for each value of $ {w}$ in  Fig.~\ref{fig:space-time-w20}.

 \begin{figure}[b!] 
\includegraphics[width=0.5\textwidth]{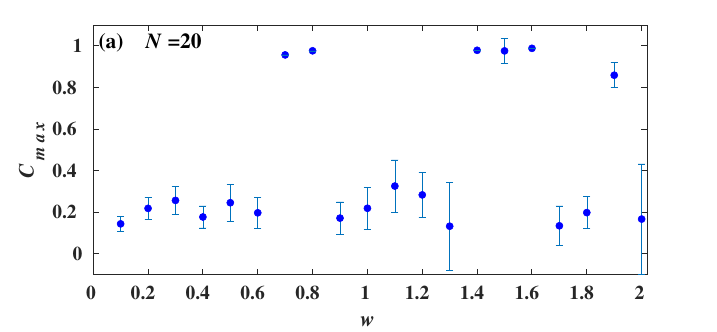} 
\includegraphics[width=0.5\textwidth]{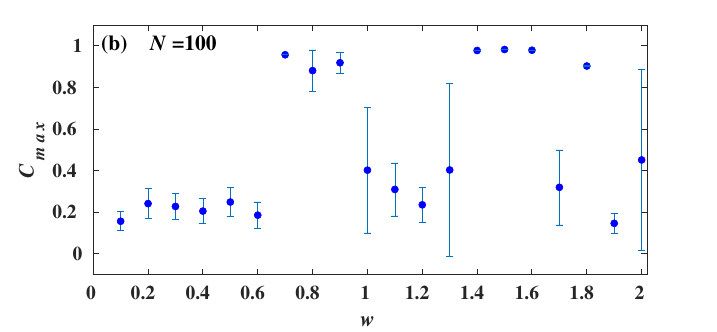}
	\caption{Maximum value of the cross-correlation between $x_1$ and $x_N$, vs. the signal's angular frequency $w$, for a chain that has (a) $N = 20$ neurons (here $D_x=0.04$), and (b) $N = 100$ neurons (here $D_x=0.1$). In both panels the signal's amplitude is $A = 0.3$. The cross-correlation is calculated over the time window $t\in [800,1000]$.  Averages (symbols) and standard deviation (bars) over 100 realizations, starting from random initial conditions, are shown.
  }
 \label{fig:CC20-100}
\end{figure}

 In the cases where the signal does not propagate (e.g., for $ {w} =0.4$), the cross-correlation between the first and last neurons is not periodic and its maximum value ($C_{max}$) is less than $0.5$. On the other hand, for $ {w} =0.7$, which is close to a resonant value,  the signal propagates reaching the end of the chain. In such case, the correlation between the first and last neurons is periodic (with period $\simeq 2\pi/ {w}$) and its maximum value is $\approx 1$, which characterizes a perfect correlation and means that the (normalized) perturbed time series $x_1$ propagated without deformation. 
 Note that, since the membrane potential is not symmetric, when $x_1$ and $x_N$ are in counter-phase the correlation in absolute value is smaller that when they are in phase.

 To summarize the outcomes obtained when varying the angular frequency, we plot  the maximum value of the cross-correlation vs. $ {w}$ in Fig.~\ref{fig:CC20-100}(a) for a network of size $N=20$. 
 The symbols  indicate the average over 100 realizations and the error bars indicate the standard deviation.
As expected, resonance occurs for frequency bands around to integer multiples of the natural oscillation frequency of the uncoupled neurons, which for the parameters used is $ {w}\simeq 0.75$. 
Let us remark that error bars can be  shorter and longer, depending on the signal frequency: large error bars occur when the frequency is near the edge of a frequency band for which the signal can propagate. In this situation, an inspection of the time series in general reveals that  in some realizations the signal reached the end of the chain (maximum cross-correlation is nearly 1), while in other realizations,   the signal did not reach the end (and the maximum cross-correlation is low), indicating a higher sensitivity to initial conditions near critical values. 

In Fig.~\ref{fig:CC20-100}(b), we show the response profile for $N=100$.  
In this case, we used a larger coupling, $D_x=0.1$, 
because, as shown in the next section, for a  fixed value of $D_x$, propagation degrades in longer chains. 
Comparing the two panels of Fig.~\ref{fig:CC20-100}, we note that the frequency bands that facilitate or prevent propagation are similar, although some differences are observed mainly in the band borders, which may require a finer adjustment of $D_x$ to achieve higher similarity.

 \begin{figure}[h!] 
\includegraphics[width=0.5\textwidth]{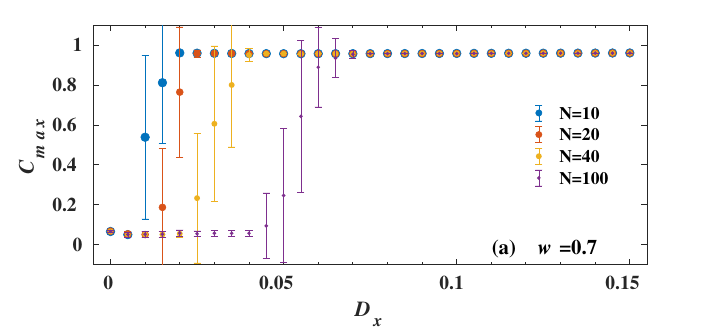}
\includegraphics[width=0.5\textwidth]{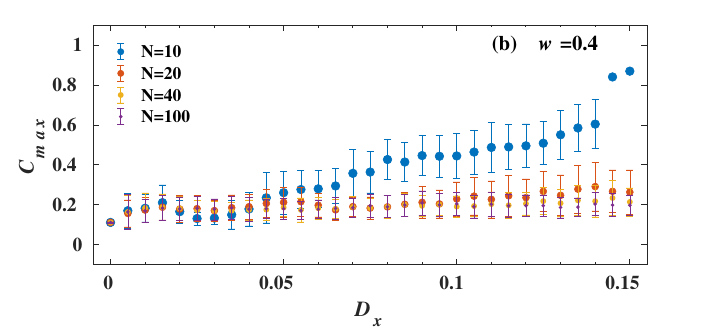} 
\caption{ 
Maximum cross-correlation between $x_1$ and $x_N$  vs. the coupling strength, $D_x$, for different sizes of the neuronal chain. The signal's parameters are $ {w}=0.7$ (a) or $ {w}=0.4$ (b), and $A=0.3$. The cross-correlation is calculated over the time window $\in [800,1000]$. Averages (symbols) and standard deviation (bars) are calculated over 100 realizations, starting from random initial conditions.
}
 \label{fig:Dx}
\end{figure}

\subsection{Effect of the coupling strength}

The effect of  the diffusive coupling strength is shown in 
Fig.~\ref{fig:Dx}, for different values of the length of the chain, $N$, and two different frequencies (resonant and non resonant).
Note that in the former case, $D_x$ must exceed a threshold value to allow propagation independently of the initial conditions and this threshold increases with $N$. Also in this case, large standard deviations (error bars) occur near a change of regime.
 Note that the values used in Fig.~\ref{fig:CC20-100}, namely,   $D_x=0.04$ (a) and 0.1 (b) are suitable choices for $N=20$ and 100, respectively.
 For the non-resonant signal, with $w=0.4$, propagation is possible only in short enough chains of strongly coupled neurons.

\subsection{Effect of the signal's amplitude}

\begin{figure}[b!] 
\includegraphics[width=0.5\textwidth]{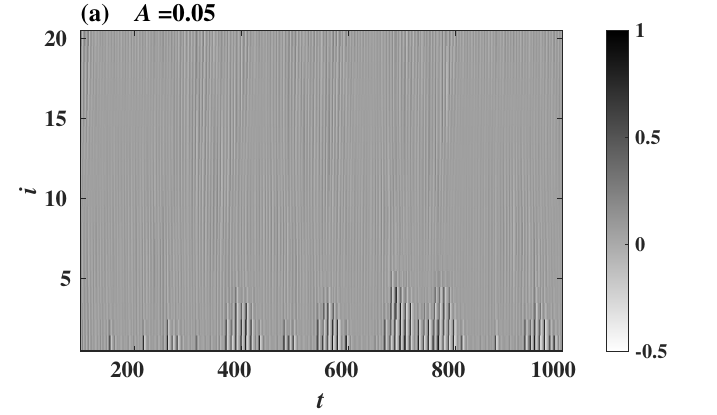}
\includegraphics[width=0.5\textwidth]{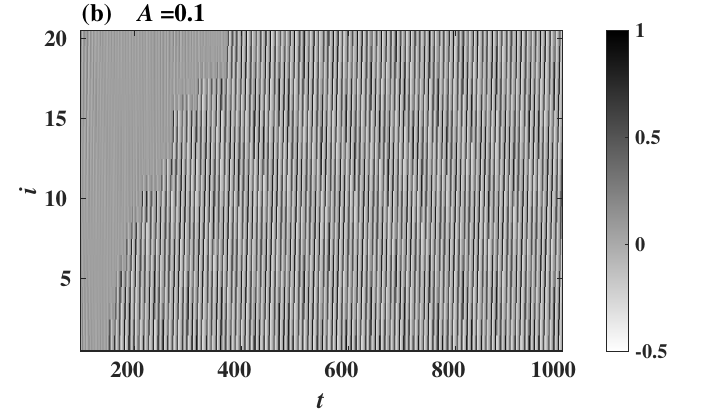}
 \caption{Raster plots of the neurons' oscillations, when the chain has $N=20$ neurons. In (a) the signal's amplitude, $A=0.05$, does not allow the signal to propagate; in (b) the amplitude is larger, $A=0.1$, and allows for signal propagation. Other parameters are $D_x=0.04$ and $ {w} = 0.7$. 
 }
 \label{fig:Amplitude}
\end{figure}

 Raster plots illustrative of the neurons' behavior for two different amplitudes, with $w=0.7$ (nearly resonant case) are presented in Fig.~\ref{fig:Amplitude}. 
Even if the frequency of the signal is close to the resonance frequency of the neurons and thus facilitates the propagation of the signal, we observe that, for small amplitude, the signal does not reach the other end.  
In panel (b), for larger amplitude, we see that the signal propagates and reaches the other end.

\begin{figure}[h!] 
\includegraphics[width=0.5\textwidth]{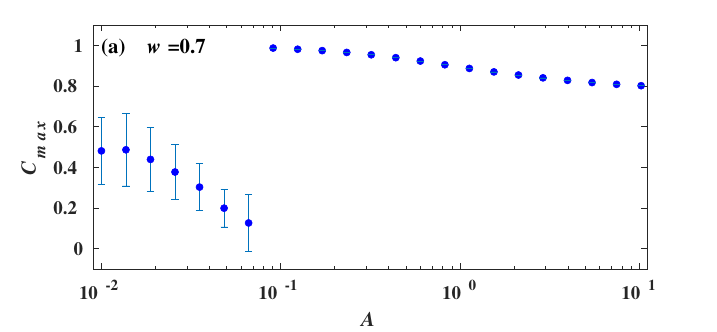} 
\includegraphics[width=0.5\textwidth]{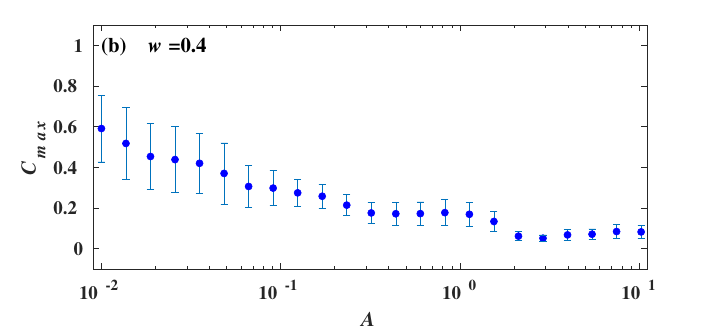}
\caption{
 Maximum value of the cross-correlation between $x_1$ and $x_N$, vs. the signal's amplitude $A$ in log scale. The signal's angular frequency is (a) $ {w} = 0.7$, and (b)  $ {w} = 0.4$. Other parameters are $N = 20$ and $D_x=0.04$. As in previous figures, the cross-correlation is calculated over the time window $\in [800,1000]$. Averages (symbols) and standard deviation (bars) over 100 realizations, starting from random initial conditions, are shown.
 }
 \label{fig:CC-A-w07}
\end{figure}

The maximum cross-correlation $C_{max}$ as a function of $A$, for $ {w}=0.7$, is shown in Fig.~\ref{fig:CC-A-w07}(a), where we see that there is a threshold value of the amplitude below which the signal does not reach the opposite end (for the parameters considered here, $A\simeq 0.08$). For both regimes (i.e., for signal's amplitudes above or below the threshold for signal propagation), the cross-correlation deteriorates with  increasing amplitude. 
That is, the network is selectively 
responsive  to excitation inputs, acting as a filter that ensures that  only significant (but not too large) inputs spread fast, which may be crucial for maintaining coherent   transmission in neuronal networks.

We also checked the effect of the amplitude when the signal's frequency is non-resonant, e.g., $w=0.4$.  The plot of $C_{max}$ vs. $A$ displayed in   Fig.~\ref{fig:CC-A-w07}(b) reveals that the signal fails to propagate,  because $C_{max}$ is small and tends to decrease with increasing $A$.

Let us remark that the results obtained in this section are consistent with the results reported in Refs.~\cite{baier1999local,baier2000human} that studied chains of calcium oscillators, also modelled by FHN diffusely coupled units.

\subsection{Injecting white noise}

Here we study the effect of white-noise added to the sinusoidal signal, and also,  the effect of injecting solely white-noise, where all frequencies are present. 

Here we generate the white noise  via  the Box-Muller method  for generating pairs of independent, standard normally distributed (Gaussian) random numbers, given two independent uniformly distributed random numbers $r_1$ and $r_2$ in the interval $(0,1)$, yielding Gaussian distributed numbers $G$ with zero mean and standard deviation $g$, through the expression $G = g  \sqrt{-2\ln(r_1)} \; \cos(2 \pi   r_2)$. 

\begin{figure}[t!]  
\includegraphics[width=0.5\textwidth]{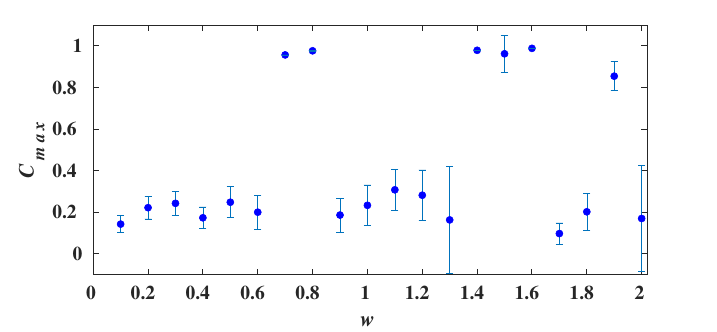} 
	\caption{ 
 Maximum cross-correlation between $x_1$ and $x_N$, vs. the signal's frequency $ {w}$, as in Fig.~\ref{fig:CC20-100}(a) with the addition of white noise with $g = 0.3$. 
The network size is $N=20$.
Averages (symbols) and standard deviation (bars) over 100 realizations, starting from random initial conditions, are shown. 
}
 \label{fig:CrN20longtime}
\end{figure}

\begin{figure}[hb!]  
  \includegraphics[width=0.5\textwidth]{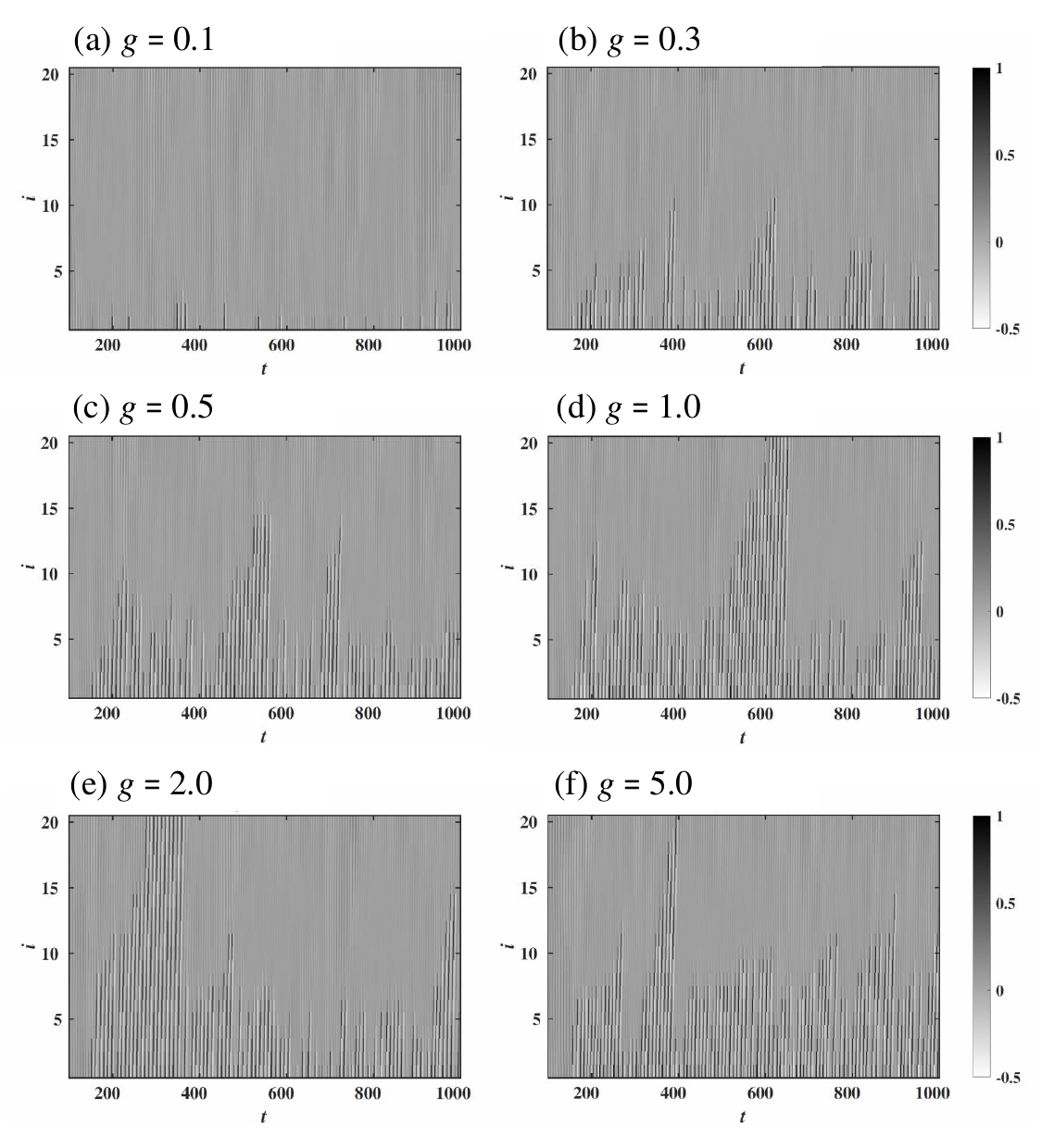}
	\caption{Spatio-temporal dynamics of the neurons when the input signal is white noise with different amplitudes of the Gaussian noise, $g\in[0,5]$. We see that strong enough noise induces spikes that can reach the other end of the chain, but the spiking dynamics does not persist. }
 \label{fig:whitenoise-g}
\end{figure}

The maximum cross-correlation as a function of $ {w}$, for the case where the injected sinusoidal signals have amplitude $A=0.3$ plus additive white noise with $g=0.3$,
is shown in Fig.~\ref{fig:CrN20longtime}.   
The plot exhibits no significant differences,  when compared to Fig.~\ref{fig:CC20-100}(a), where noise is absent. 
Similar profile is also observed for $g=2.0$ (not shown).
Moreover, for $ {w}=0.7$, we plotted  the maximal correlation as a function of $g\in[0,2]$ yielding a very flat profile at a level $C_{max}\simeq 0.96$ (not shown). 
These comparisons indicate that, at least for the values of the parameters within the investigated ranges, the results are robust under the addition of white noise. 
 
\begin{figure}[h!]  
\includegraphics[width=0.45\textwidth]{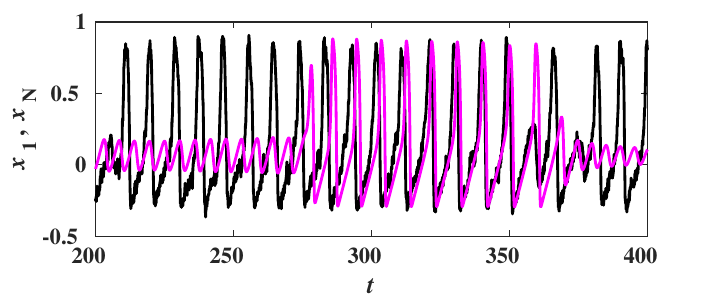}
	\caption{Time series of the membrane potentials of the end neurons, $x_1$ (black) and $x_N$ (magenta), when the input signal injected in the first neuron is Gaussian white noise with  amplitude $g=2.0$.  Other parameters are $N=20$, $D_x=0.04$.}
 \label{fig:twhitenoise-g2}
\end{figure}

 We inspected the effects of white-noise signals injected alone (without regular oscillations) into the system (with $A=0$). We note, in Fig.~\ref{fig:whitenoise-g}, that even for high amplitudes of the noise, persistent propagation does not occur. 
Although all frequencies are contained in the white noise, in particular resonant ones, their contribution does not have enough amplitude for propagation. 
However, note that, despite not reaching the end of the chain, the fronts tend to propagate temporarily. 
In Fig.~\ref{fig:twhitenoise-g2} we show segments of the time-series of $x_1$ and $x_N$ for which transient propagation has occurred. Also note that the induced spikes that reach the last neuron have a shape similar to that of the sinusoidal input with fundamental resonant frequency, as in Fig.~\ref{fig:space-time-w20}(a). 

\subsection{Subcritical neurons}

Here we analyze the propagation in a linear chain of neurons that have a slightly subcritical external current, e.g., $I_{ext}=0.05$. 
In this case, when adding the sinusoidal signal to a single (isolated) FHN neuron, after a transient the neuron evolves in a limit cycle, while without the sinusoidal signal, the neuron remains in a stable steady state (a focus) in the absence of external perturbations.

\begin{figure}[ht!]  
\includegraphics[width=0.5\textwidth]{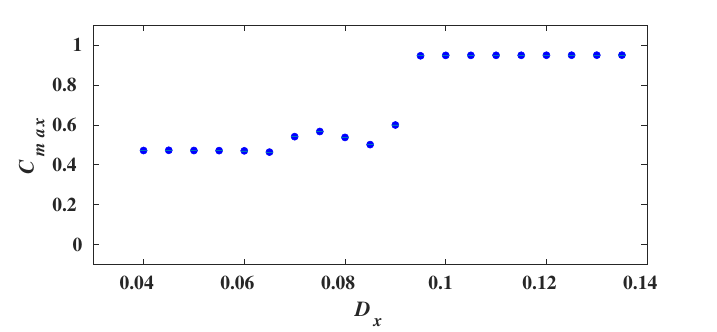} 
	\caption{ 
 Maximum cross-correlation between $x_1$ and $x_N$, vs. coupling strength, $D_x$, for a linear chain of $N=20$ subcritical neurons with $I_{ext}=0.05$. The signal's parameters are $ {w}=0.7$ and $A=0.3$. As in previous figures, $C$ is calculated over the time window $\in [800,1000]$.
Averages (symbols) and standard deviation (bars) were computed over 100 realizations. }
 \label{fig:subcritical-Dx}
\end{figure}
\begin{figure}[hb!] 
\includegraphics[width=0.47\textwidth]{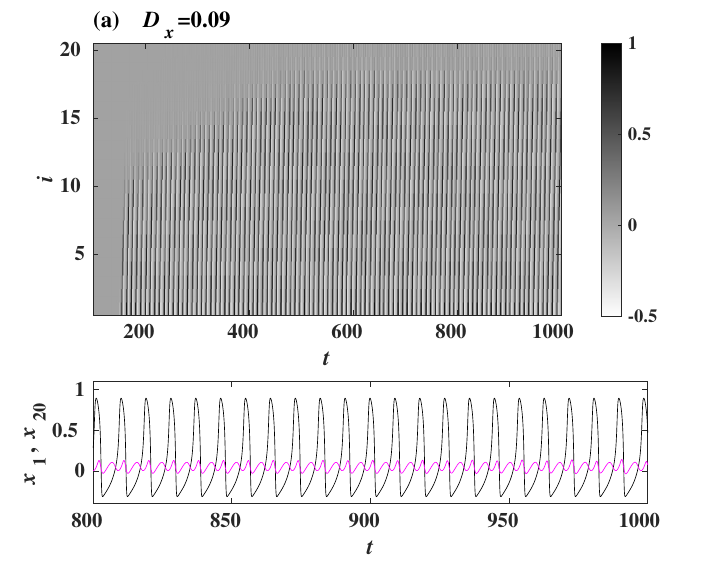}
\includegraphics[width=0.47\textwidth]{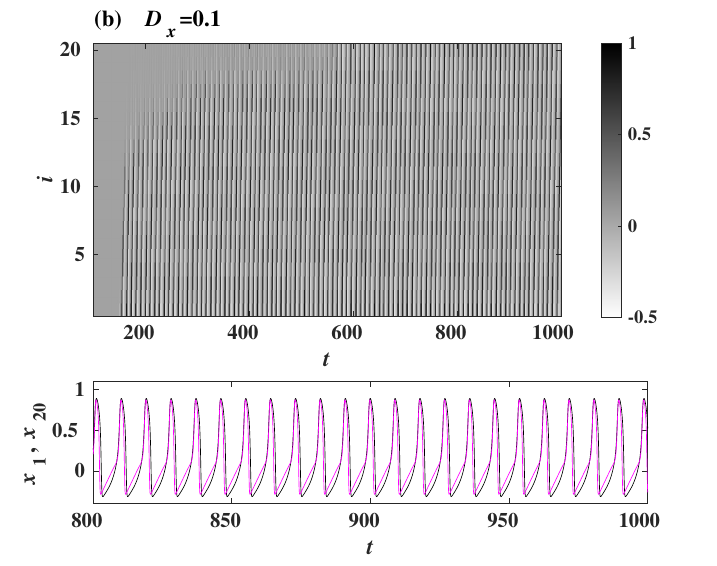}
 \caption{Raster plots of the neurons' oscillations (a) and timeseries of the end neurons (b), when the chain has $N=20$ subcritical neurons and the signal's parameters are $A=0.3$ and $ {w}=0.7$. The coupling coefficient is (a) $D_x=0.08$ and (b) $D_x=0.09$.
 }
 \label{fig:spacetime-Dx}
\end{figure}

For a chain of 20 subcritical neurons, with the same coupling coefficient as before, $D_x=0.04$, signals event with frequencies close to the resonant ones present only moderate maximal cross-correlation. 
This motivated us to study the effect of the coupling coefficient, $D_x$ in this subcritical case. 
In fact, we observe  a threshold value of $D_x$, above which propagation can indeed occur, as shown in Fig.~\ref{fig:subcritical-Dx}. 
In such case,  neurons that are in an inactive state can be excited with the sinusoidal input stimulus to propagate information, if the coupling strength is high enough.
In Fig.~\ref{fig:spacetime-Dx}, we show the raster plots of the first and last neurons, $x_1$ and $x_N$, for $D_x$
below and above the critical point. In the first case, panel (a), only small oscillations are observed in $x_N$, while
in the second case, panel (b), the spikes in $x_N$ are perfectly correlated with those in $x_1$.

 \subsection{Role of the waiting time}
 
\begin{figure}[h!] 
\includegraphics[width=0.5\textwidth]{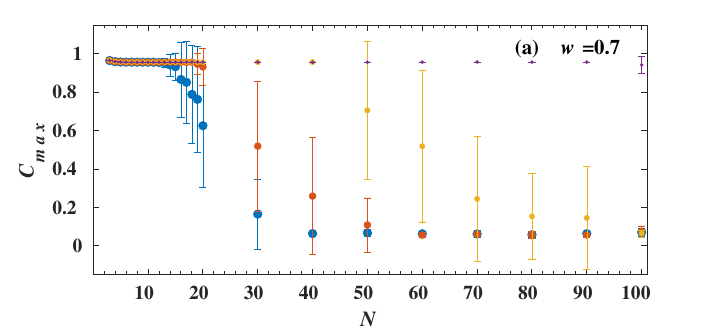}  
\includegraphics[width=0.5\textwidth]{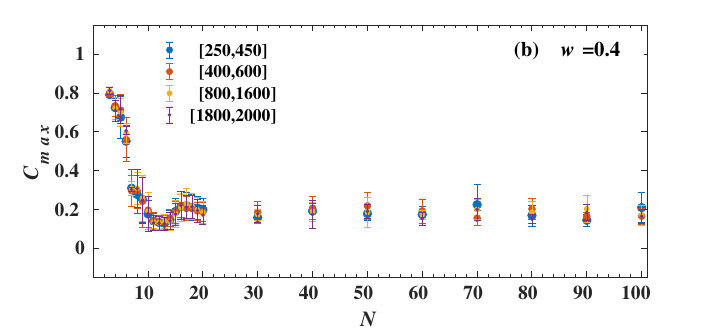} 
\caption{Maximum cross-correlation between $x_1$ and $x_N$  vs. number of neurons in the chain $N$, for time windows $[t_0,t_0+\Delta t]$ indicated in the legend, for two values of $w$: 0.7 (a) and 0.4 (b). The coupling is $D_x=0.04$ and the amplitude of the signal is $A=0.3$.
}
\label{fig:window}
\end{figure}

We conclude by analyzing the effect of the waiting time, i.e., the start of the analysis window, $[t_0,t_0+\Delta t]$, used to quantify the propagation of the signal by calculating the cross-correlation between $x_1$ and $x_N$. As explained before, we keep $\Delta t=200$ constant because it allows including in the analysis window several periods of the signal, for all the range of frequencies considered. Fig.~\ref{fig:window} displays the maximum of the cross-correlation vs. the length of the chain, $N$, for various values of $t_0$. 
We see that a waiting time $t_0=800$ is appropriate if the chain has up to $N\simeq 20$, however, longer waiting times are required for longer chains. For $N=100$, $t_0$ must be longer than 2000.\\

\section{Conclusion and outlook}
 
We have studied the propagation of simple sinusoidal signals (which can also be noisy) in excitable media, using a toy model that consists of a linear chain of diffusively coupled FHN neurons and injecting the signal at one end of the chain. 
Our study continues previous efforts to understand the propagation of signals in linear chains of neurons as excitable media~\cite{baier1999local,baier2000human,chernihovskyi2005}.

We were interested in detecting if the signal injected in the membrane potential of the first neuron arrives to the neuron at the other end of the chain. 
To quantify this propagation, we introduced the cross-correlation coefficient $C$ between the membrane potentials of both neurons as a function of the time lag. We have shown that its maximum value $C_{max}$ represents an efficient measure to assess whether the injected signal reaches the end neuron. 
We have shown that this procedure allows to detect threshold values of the parameters, delimiting regimes with or without propagation.
We tested this method by varying several relevant parameters of the signal, such as amplitude $A$, frequency $ {w}$, level of added noise $g$. 
Although we kept the local dynamics fixed, using the same value of the neuron parameters used in previous works~\cite{baier1999local,baier2000human,chernihovskyi2005}, 
we varied the parameters that characterize the excitable system, such as the diffusive coupling coefficient $D_x$ between neurons and chain size $N$, enriching the knowledge about the nonlinear dynamics of this excitable type of system.

In this paper, we have tested the method using simple (sinusoidal)  input signals.  An interesting next step is to study more complex signals, for example, signals that are the superposition of several sinusoidals with different frequencies, or a sinusoidal signal whose amplitude and frequency vary over time. 
  In addition, the method could be extended to real-world signals, such as physiological, climatic, or financial data.  
 This is motivated by the potential practical application of the chain as a signal processing tool, as it could be used to differentiate periodic signals with different frequency band contents. 
It would also be interesting to explore other coupling schemes, such as pulse-based couplings, as well as to investigate networks beyond linear chains and alternative neuron models\cite{Izhikevich2007,HRmodel,budzinski2020,birhythmic2023,Messee2023,budzinski2023}. 
Another interesting perspective for future work is to quantify the propagation of the signal using nonlinear quantifiers, in particular, instead of correlating the whole time series of the membrane potential of the two neurons at the end of the linear chain, one could correlate the timing of the spikes fired by those neurons, by using spike-based synchronization measures~\cite{event_sync}.

\section*{Acknowledgments}

L.M.G.  acknowledges the Organization for Women Scientists for Developing World (OWSD Postgraduate Fellowship, Grant No. 3240318616) and Swedish International Development Cooperation Agency (SIDA) for financial support. C.A. acknowledges partial financial support from Brazilian agencies Conselho Nacional de Desenvolvimento Científico 
(CNPq, 311435/2020-3),  
Fundação Carlos Chagas Filho de Amparo \`a Pesquisa do Estado do Rio de Janeiro (FAPERJ, CNE E-26/204.130/2024)
and Coordenação de Aperfeiçoamento de Pessoal de Nível Superior 
(CAPES, code 001). C. M. acknowledges partial financial support  of Agència de Gestió d’Ajuts Universitaris i de Recerca (2021 SGR 00606), the Institució
Catalana de Recerca i Estudis Avançats (Academia), the Ministerio de Ciencia, Innovación y Universidades (PID2021-123994NB-C21) and the European Office of Aerospace Research and Development (FA8655-24-1-7022).

\end{document}